\pdfoutput=1
\documentclass[12pt,a4paper]{article}

\usepackage{ifthen} 
\newboolean{pdflatex}
\setboolean{pdflatex}{true} 

\newboolean{articletitles}
\setboolean{articletitles}{true} 

\newboolean{uprightparticles}
\setboolean{uprightparticles}{false} 

\def\paperauthors{LHCb collaboration} 
\def\paperasciititle{Precise determination of the B0s–antiB0s oscillation frequency} 
\def\papertitle{Precise determination \\
of the $\Bs$--$\Bsb$ oscillation frequency}
\def\paperkeywords{{High Energy Physics}, {LHCb}} 
\def\papercopyright{\the\year\ CERN for the benefit of the LHCb collaboration} 
\def\paperlicence{CC BY 4.0 licence}
\def\paperlicenceurl{https://creativecommons.org/licenses/by/4.0/}

\usepackage[top=1in, bottom=1.25in, left=1in, right=1in]{geometry}

\usepackage{microtype}
\usepackage{lineno}  
\usepackage{xspace} 
\usepackage{caption} 

\usepackage{graphicx}  
\usepackage{color}
\usepackage{colortbl}
\graphicspath{{./figs/}} 

\usepackage{amsmath} 
\usepackage{amssymb}
\usepackage{amsfonts}
\usepackage{upgreek} 

\newcommand*\patchAmsMathEnvironmentForLineno[1]{%
\expandafter\let\csname old#1\expandafter\endcsname\csname #1\endcsname
\expandafter\let\csname oldend#1\expandafter\endcsname\csname
end#1\endcsname
 \renewenvironment{#1}%
   {\linenomath\csname old#1\endcsname}%
   {\csname oldend#1\endcsname\endlinenomath}%
}
\newcommand*\patchBothAmsMathEnvironmentsForLineno[1]{%
  \patchAmsMathEnvironmentForLineno{#1}%
  \patchAmsMathEnvironmentForLineno{#1*}%
}
\AtBeginDocument{%
\patchBothAmsMathEnvironmentsForLineno{equation}%
\patchBothAmsMathEnvironmentsForLineno{align}%
\patchBothAmsMathEnvironmentsForLineno{flalign}%
\patchBothAmsMathEnvironmentsForLineno{alignat}%
\patchBothAmsMathEnvironmentsForLineno{gather}%
\patchBothAmsMathEnvironmentsForLineno{multline}%
\patchBothAmsMathEnvironmentsForLineno{eqnarray}%
}

\usepackage{hyperxmp}

\usepackage[pdftex,
            pdfauthor={\paperauthors},
            pdftitle={\paperasciititle},
            pdfkeywords={\paperkeywords},
            pdfcopyright={Copyright (C) \papercopyright},
            pdflicenseurl={\paperlicenceurl}]{hyperref}

\usepackage[colorinlistoftodos,textsize=scriptsize]{todonotes}

\usepackage[bottom,flushmargin,hang,multiple]{footmisc}

\usepackage[all]{hypcap} 

\def\lhcb   {\mbox{LHCb}\xspace}

\def\MagUp {\mbox{\em Mag\kern -0.05em Up}\xspace}

\ifthenelse{\boolean{uprightparticles}}%
{

 \def\Ppi         {\ensuremath{\uppi}\xspace}

 \def\Ppsi        {\ensuremath{\uppsi}\xspace}

 \def\PDelta      {\ensuremath{\Delta}\xspace}                 
 \def\PXi         {\ensuremath{\Xi}\xspace}                 
 \def\PLambda     {\ensuremath{\Lambda}\xspace}                 
 \def\PSigma      {\ensuremath{\Sigma}\xspace}                 
 \def\POmega      {\ensuremath{\Omega}\xspace}                 
 \def\PUpsilon    {\ensuremath{\Upsilon}\xspace}

 \def\PB      {\ensuremath{\mathrm{B}}\xspace}                 
                  
 \def\PD      {\ensuremath{\mathrm{D}}\xspace}

 \def\PJ      {\ensuremath{\mathrm{J}}\xspace}                 
 \def\PK      {\ensuremath{\mathrm{K}}\xspace}

 \def\Pb      {\ensuremath{\mathrm{b}}\xspace}                 
 \def\Pc      {\ensuremath{\mathrm{c}}\xspace}                 
 \def\Pd      {\ensuremath{\mathrm{d}}\xspace}

 \def\Pi      {\ensuremath{\mathrm{i}}\xspace}

 \def\Ps      {\ensuremath{\mathrm{s}}\xspace}

 \def\thebaroffset{0.0em}
}
{

 \def\Ppi         {\ensuremath{\pi}\xspace}

 \def\Ppsi        {\ensuremath{\psi}\xspace}                 
                  
 \mathchardef\PDelta="7101
 \mathchardef\PXi="7104
 \mathchardef\PLambda="7103
 \mathchardef\PSigma="7106
 \mathchardef\POmega="710A
 \mathchardef\PUpsilon="7107
                  
 \def\PB      {\ensuremath{B}\xspace}                 
                  
 \def\PD      {\ensuremath{D}\xspace}

 \def\PJ      {\ensuremath{J}\xspace}                 
 \def\PK      {\ensuremath{K}\xspace}

 \def\Pb      {\ensuremath{b}\xspace}                 
 \def\Pc      {\ensuremath{c}\xspace}                 
 \def\Pd      {\ensuremath{d}\xspace}

 \def\Pi      {\ensuremath{i}\xspace}

 \def\Ps      {\ensuremath{s}\xspace}

 \def\thebaroffset{0.18em}
}
\newcommand{\offsetoverline}[2][\thebaroffset]{\kern #1\overline{\kern -#1 #2}}%

\makeatletter
\ifcase \@ptsize \relax
  \newcommand{\miniscule}{\@setfontsize\miniscule{4}{5}}
\or
  \newcommand{\miniscule}{\@setfontsize\miniscule{5}{6}}
\or
  \newcommand{\miniscule}{\@setfontsize\miniscule{5}{6}}
\fi
\makeatother

\DeclareRobustCommand{\optbar}[1]{\shortstack{{\miniscule (\rule[.5ex]{1.25em}{.18mm})}
  \\ [-.7ex] $#1$}}

\def\dquark    {{\ensuremath{\Pd}}\xspace}

\def\squark    {{\ensuremath{\Ps}}\xspace}

\def\cquark    {{\ensuremath{\Pc}}\xspace}

\def\bquark    {{\ensuremath{\Pb}}\xspace}

\def\pion   {{\ensuremath{\Ppi}}\xspace}

\def\pip    {{\ensuremath{\pion^+}}\xspace}
\def\pim    {{\ensuremath{\pion^-}}\xspace}
\def\pipm   {{\ensuremath{\pion^\pm}}\xspace}

\def\kaon    {{\ensuremath{\PK}}\xspace}

\def\KorKbar {\kern \thebaroffset\optbar{\kern -\thebaroffset \PK}{}\xspace}

\def\Kp      {{\ensuremath{\kaon^+}}\xspace}
\def\Km      {{\ensuremath{\kaon^-}}\xspace}

\def\D       {{\ensuremath{\PD}}\xspace}

\def\DorDbar {\kern \thebaroffset\optbar{\kern -\thebaroffset \PD}\xspace}

\def\Dp      {{\ensuremath{\D^+}}\xspace}
\def\Dm      {{\ensuremath{\D^-}}\xspace}

\def\DpDm    {\ensuremath{\Dp {\kern -0.16em \Dm}}\xspace}

\def\Ds      {{\ensuremath{\D^+_\squark}}\xspace}
\def\Dsp     {{\ensuremath{\D^+_\squark}}\xspace}
\def\Dsm     {{\ensuremath{\D^-_\squark}}\xspace}

\def\Dssm    {{\ensuremath{\D^{*-}_\squark}}\xspace}

\def\B       {{\ensuremath{\PB}}\xspace}
\def\Bbar    {{\ensuremath{\offsetoverline{\PB}}}\xspace}

\def\BorBbar {\kern \thebaroffset\optbar{\kern -\thebaroffset \PB}\xspace}
\def\Bz      {{\ensuremath{\B^0}}\xspace}

\def\Bd      {{\ensuremath{\B^0}}\xspace}
\def\Bdb     {{\ensuremath{\Bbar{}^0}}\xspace}
\def\BdorBdbar {\kern \thebaroffset\optbar{\kern -\thebaroffset \Bd}\xspace}
\def\Bu      {{\ensuremath{\B^+}}\xspace}

\def\Bp      {{\ensuremath{\Bu}}\xspace}

\def\Bs      {{\ensuremath{\B^0_\squark}}\xspace}
\def\Bsb     {{\ensuremath{\Bbar{}^0_\squark}}\xspace}
\def\BsorBsbar {\kern \thebaroffset\optbar{\kern -\thebaroffset \Bs}\xspace}

\def\jpsi     {{\ensuremath{{\PJ\mskip -3mu/\mskip -2mu\Ppsi}}}\xspace}

\def\Y#1S{\ensuremath{\PUpsilon{(#1S)}}\xspace}

\def\Lbar        {{\ensuremath{\offsetoverline{\PLambda}}}\xspace}
\def\LorLbar     {\kern \thebaroffset\optbar{\kern -\thebaroffset \PLambda}\xspace}

\def\Lcbar       {{\ensuremath{\Lbar{}^-_\cquark}}\xspace}

\def\Lbbar        {{\ensuremath{\Lbar{}^0_\bquark}}\xspace}

\newcommand{\decay}[2]{\ensuremath{#1\!\to #2}\xspace} 

\def\to                 {\ensuremath{\rightarrow}\xspace}

\newcommand{\dms}{{\ensuremath{\Delta m_{\squark}}}\xspace}
\newcommand{\dmd}{{\ensuremath{\Delta m_{\dquark}}}\xspace}

\def\AT#1     {\ensuremath{A_{\mathrm{T}}^{#1}}\xspace}           

\def\C#1      {\ensuremath{\mathcal{C}_{#1}}\xspace}                       
\def\Cp#1     {\ensuremath{\mathcal{C}_{#1}^{'}}\xspace}                    
\def\Ceff#1   {\ensuremath{\mathcal{C}_{#1}^{\mathrm{(eff)}}}\xspace}        
\def\Cpeff#1  {\ensuremath{\mathcal{C}_{#1}^{'\mathrm{(eff)}}}\xspace}       
\def\Ope#1    {\ensuremath{\mathcal{O}_{#1}}\xspace}                       
\def\Opep#1   {\ensuremath{\mathcal{O}_{#1}^{'}}\xspace}                    

\newcommand{\nospaceunit}[1]{\ensuremath{\text{#1}}}       
\newcommand{\aunit}[1]{\ensuremath{\text{\,#1}}}       

\newcommand{\tev}{\aunit{Te\kern -0.1em V}\xspace}
\newcommand{\gev}{\aunit{Ge\kern -0.1em V}\xspace}
\newcommand{\mev}{\aunit{Me\kern -0.1em V}\xspace}
\newcommand{\kev}{\aunit{ke\kern -0.1em V}\xspace}
\newcommand{\ev}{\aunit{e\kern -0.1em V}\xspace}
 
\newcommand{\mevc}{\ensuremath{\aunit{Me\kern -0.1em V\!/}c}\xspace}
\newcommand{\gevc}{\ensuremath{\aunit{Ge\kern -0.1em V\!/}c}\xspace}
\newcommand{\mevcc}{\ensuremath{\aunit{Me\kern -0.1em V\!/}c^2}\xspace}
\newcommand{\gevcc}{\ensuremath{\aunit{Ge\kern -0.1em V\!/}c^2}\xspace}

\def\m    {\aunit{m}\xspace}

\def\cm   {\aunit{cm}\xspace}

\def\mum  {\ensuremath{\,\upmu\nospaceunit{m}}\xspace}

\def\fb   {\ensuremath{\aunit{fb}}\xspace}
\def\invfb   {\ensuremath{\fb^{-1}}\xspace}

\def\ps   {\ensuremath{\aunit{ps}}\xspace}

\newcommand{\stat}{\aunit{(stat)}\xspace}
\newcommand{\syst}{\aunit{(syst)}\xspace}

\def\gsim{{~\raise.15em\hbox{$>$}\kern-.85em
          \lower.35em\hbox{$\sim$}~}\xspace}
\def\lsim{{~\raise.15em\hbox{$<$}\kern-.85em
          \lower.35em\hbox{$\sim$}~}\xspace}

\def\pt         {\ensuremath{p_{\mathrm{T}}}\xspace}

\def\ptot       {\ensuremath{p}\xspace}

\def\mrad{\aunit{mrad}\xspace}

\def\evtgen     {\mbox{\textsc{EvtGen}}\xspace}

\def\geant      {\mbox{\textsc{Geant4}}\xspace}

\def\photos     {\mbox{\textsc{Photos}}\xspace}

\def\pythia     {\mbox{\textsc{Pythia}}\xspace}

\def\tell1  {TELL1\xspace}
\def\ukl1   {UKL1\xspace}

\newcommand{\ie}{\mbox{\itshape i.e.}\xspace}

\def\combinationAll{17.7656 \pm 0.0057}
\def\combinationHadronic{17.7666 \pm 0.0057}
\def\resultstat{17.7683 \pm 0.0051}
\def\resultsyst{0.0032}
\def\result{\resultstat \stat \pm \resultsyst \syst}

\renewcommand{\Ds}       {\texorpdfstring{\ensuremath{D_s^-}}{Ds}\xspace}

\newcommand{\BdDPi}    {\texorpdfstring{\decay{\Bz}{\Dm \pip}}{}}

\newcommand{\BdDsPi}   {\texorpdfstring{\decay{\Bz}{\Dsm \pip}}{}}

\newcommand{\BsDsK}    {\texorpdfstring{\decay{B^0_s}{D_s^\mp K^\pm}}{}}
\newcommand{\BsDsPi}   {\texorpdfstring{\decay{B^0_s}{D_s^- \pi^+}}{}}

\newcommand{\BsbDsPi}   {\texorpdfstring{\decay{\Bsb}{D_s^- \pi^+}}{}}
\newcommand{\BsbDsPim}   {\texorpdfstring{\decay{\Bsb}{D_s^+ \pi^-}}{}}

\newcommand{\BsDsstPi} {\decay{\Bs}{\Dssm \pip}}

\newcommand{\DsKKPi}   {\decay{\Dsm}{\Km\Kp\pim}}

\newcommand{\DsPiPiPi} {\decay{\Dsm}{\pim\pip\pim}}

\newcommand{\LbLcPi}   {\decay{\Lbbar}{\Lcbar \pip}}

\usepackage{cite} 
\usepackage{mciteplus}

\usepackage{longtable} 

\begin{document}

\renewcommand{\thefootnote}{\fnsymbol{footnote}}
\setcounter{footnote}{1}


\begin{titlepage}
\pagenumbering{roman}

\vspace*{-1.5cm}
\centerline{\large EUROPEAN ORGANIZATION FOR NUCLEAR RESEARCH (CERN)}
\vspace*{1.5cm}
\noindent
\begin{tabular*}{\linewidth}{lc@{\extracolsep{\fill}}r@{\extracolsep{0pt}}}
\ifthenelse{\boolean{pdflatex}}
{\vspace*{-1.5cm}\mbox{\!\!\!\includegraphics[width=.14\textwidth]{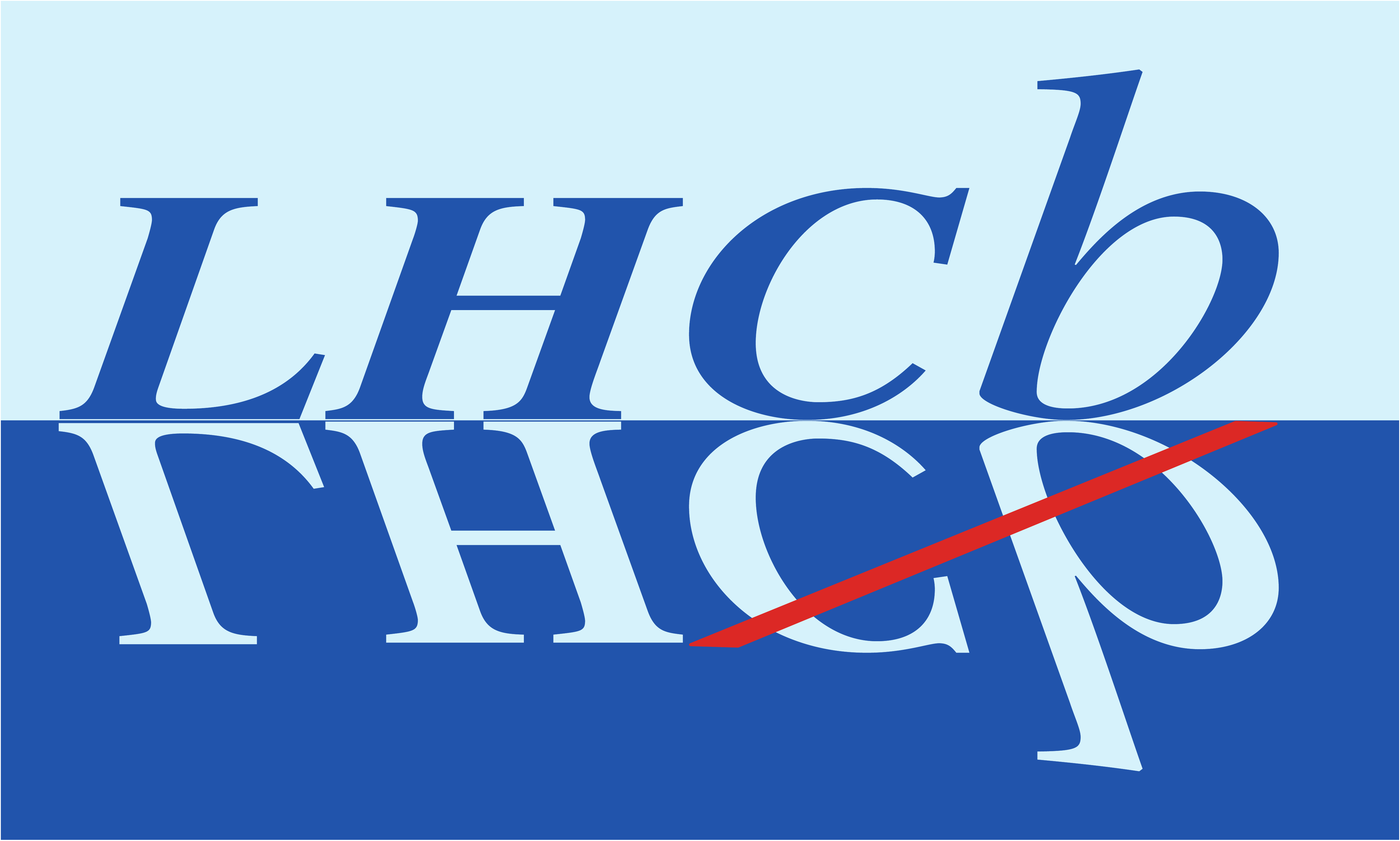}} & &}%
{\vspace*{-1.2cm}\mbox{\!\!\!\includegraphics[width=.12\textwidth]{figs/lhcb-logo.eps}} & &}%
\\
 & & CERN-EP-2021-047 \\  
 & & LHCb-PAPER-2021-005 \\  
 & & \today \\ 
 & & \\
\end{tabular*}

\vspace*{4.0cm}

{\normalfont\bfseries\boldmath\huge
\begin{center}
  \papertitle 
\end{center}
}

\vspace*{2.0cm}

\begin{center}
\paperauthors\footnote{Authors are listed at the end of this paper.}
\end{center}

\vspace{\fill}

\begin{abstract}
  \noindent
  Mesons comprising a beauty quark and a strange quark can oscillate between particle
  (\Bs) and antiparticle (\Bsb) flavour eigenstates, with a frequency given by the mass 
  difference between heavy and light mass eigenstates, \dms.
  Here we present a measurement of \dms using \BsDsPi decays produced in proton-proton 
  collisions collected with the LHCb detector at the Large Hadron Collider.
  The oscillation frequency is found to be 
  $\dms = \resultstat \pm \resultsyst \ps^{-1}$,
  where the first uncertainty is statistical and the second systematic.
  This measurement improves upon the current \dms precision by a factor of two.
  We combine this result with previous LHCb measurements to determine
  $\dms = \combinationAll \ps^{-1}$, which is the legacy measurement of the original 
  LHCb detector.
  
\end{abstract}

\vspace*{2.0cm}

\begin{center}
  Published in Nature Physics 18, (2022) 1
\end{center}

\vspace{\fill}

{\footnotesize 
\centerline{\copyright~\papercopyright. \href{\paperlicenceurl}{\paperlicence}.}}
\vspace*{2mm}

\end{titlepage}


\newpage
\setcounter{page}{2}
\mbox{~}
%
%
%
%


\renewcommand{\thefootnote}{\arabic{footnote}}
\setcounter{footnote}{0}

\cleardoublepage


\pagestyle{plain} 
\setcounter{page}{1}
\pagenumbering{arabic}



Neutral mesons with strange, charm or beauty quantum numbers can mix with their antiparticles, as these quantum numbers are not conserved by the weak interaction. The neutral meson comprising an antibeauty quark and a strange quark, the $\Bs$ meson, and its antiparticle, the $\Bsb$ meson, are one such example. In the $\Bs$--$\Bsb$ system, the observed particle and antiparticle states are linear combinations of the heavy (H) and light (L)
mass eigenstates. The mass eigenstates have masses $m_{\rm H}$ and $m_{\rm L}$ and decay widths $\Gamma_{\rm H}$ and
$\Gamma_{\rm L}$ \cite{DeltaGamma}.
As a consequence, the $\Bs$--$\Bsb$ system oscillates with a frequency given by the mass difference, $\dms = m_{\rm H} - m_{\rm L}$.
This oscillation frequency is an important parameter of the Standard Model of particle
physics.
In combination with the $\Bd$--$\Bdb$ oscillation frequency, \dmd, it 
provides a powerful constraint on the Cabibbo–Kobayashi–Maskawa quark-mixing
matrix~\cite{Cabibbo:1963yz,Kobayashi:1973fv,UTfit-UT,CKMfitter2015,1902.08191}.
A precise measurement of $\dms$ is also required to reduce the systematic uncertainty associated 
with measurements of matter-antimatter differences in the $\Bs$--$\Bsb$ system~\cite{LHCb-PAPER-2017-047}.

In this paper, we present a measurement of \dms using \Bs mesons that decay to a charmed-strange \Ds meson and a pion, \BsDsPi, and the decays with opposite charge, \BsbDsPim. We refer to both charge combinations as \BsDsPi throughout the paper, and similarly for decays of the \Ds meson.
The measurement is performed using data collected between 2015 and 2018, denoted Run 2 of the Large Hadron 
Collider (LHC), corresponding to an integrated luminosity of $6\invfb$ of proton-proton ($pp$) collisions 
at a centre-of-mass energy of $13\tev$.

The first measurement in which the significance of the observed
$\Bs$--$\Bsb$ oscillation signal exceed five standard deviations was obtained 
by the CDF collaboration~\cite{CDFDeltaMs}.
More recently, the LHCb collaboration has performed several measurements of \dms using data collected at 
the LHC: a measurement using \BsDsPi decays~\cite{LHCb-PAPER-2013-006}; 
two measurements using $\Bs \rightarrow J/\psi K^+ K^-$ decays~\cite{LHCb-PAPER-2014-059,LHCb-PAPER-2019-013};
and a measurement using $B_s^0 \to D_s^{\mp} \pi^{\pm} \pi^{\pm}\pi^{\mp}$ decays~\cite{LHCb-PAPER-2020-030}. 
Theoretical predictions for \dms are
available~\cite{Phys.Rev.D93.2016.113016,1812.08791,1907.01025,1806.00253,1904.00940,1902.08191},
with the most precise prediction in Ref.~\cite{DeltaMsTheoUpdate}.
The prediction is consistent with but significantly less precise than existing experimental results.

The \BsDsPi decay-time distribution, in the absence of detector effects, can be written as
\begin{equation}
P(t) \sim
    e^{-\Gamma_s t}
    \left[
        \cosh\left(\frac{\Delta\Gamma_s t}{2}\right)
        + C \cdot \cos(\Delta m_s t)
    \right]\,,
\label{eq:time_eq}
\end{equation}
where $\Gamma_s=(\Gamma_{\rm H}+\Gamma_{\rm L})/2$ is the \Bs
meson decay width and $\Delta\Gamma_s=\Gamma_{\rm H}-\Gamma_{\rm L}$ is the decay-width
difference between the heavy and light mass eigenstates. The parameter $C$ takes the value $C = 1$ for 
unmixed decays, \ie \BsDsPi, and $C = -1$ for decays in which the initially produced meson mixed into its antiparticle before decaying, 
\ie $\Bs\to\BsbDsPim$. The mixed decay is referred to as 
$\BsbDsPi$ throughout the paper. The mass difference \dms corresponds to a frequency in natural units, and 
is measured in inverse picoseconds. 

The \lhcb detector~\cite{LHCb-DP-2008-001,LHCb-DP-2014-002} is designed to study the decays of 
beauty and charm hadrons produced in $pp$ collisions at the LHC.
It instruments a region around the proton beam axis, covering the polar angles between 10 and 250\mrad, in which approximately a quarter of the \bquark-hadron decay products are fully contained.
The detector includes a high-precision tracking system with a dipole magnet, providing measurements of the momentum and decay-vertex position of particles.
Different types of charged particles are distinguished using information from two ring-imaging Cherenkov detectors, a calorimeter and a muon system.

Simulated samples of \BsDsPi decays and data control samples are used to verify the analysis 
procedure and to study systematic effects.
The simulation provides a detailed model of the experimental conditions, including the $pp$ 
collision, the decays of the particles produced, their final-state radiation and the response 
of the detector.
Simulated samples are corrected for residual differences in relevant kinematic distributions to
improve the agreement with data.
The software used is described in
Refs.~\cite{Sjostrand:2007gs,LHCb-PROC-2010-056,Lange:2001uf,Allison:2006ve, *Agostinelli:2002hh,LHCb-PROC-2011-006,davidson2015photos}.

The \Bs mesons travel a macroscopic distance at LHC energies (on average 1 cm) before decaying and are significantly heavier than most other particles produced directly in $pp$ collisions. Thus their decay products have significant displacement relative to the $pp$ collision point, and a larger momentum transverse to the beam axis, compared to other particles. The candidate selection exploits these fundamental properties.
Two fast real-time selections use partial detector information to reject LHC bunch crossings likely to be incompatible with the presence of the signal, before a third selection uses fully aligned and calibrated data in real time to reconstruct and select topologies consistent with the signal~\cite{LHCb-DP-2019-001}. Selected collisions are recorded to permanent storage.
All but the first real-time selection are based on multivariate classifiers. 
Two subsequent selections fully reconstruct the decays with the $\Ds$ meson reconstructed in both $\Km\Kp\pim$ and $\pim\pip\pim$ final states.
After the real-time stages, the initial `offline' selection is based on a data range in track kinematic 
quantities and displacement relative to the $pp$ collision point that favours signal, followed by a multivariate
classifier trained on properties of the full signal decay. 
These selections sequentially improve the signal purity of the sample to the final value of 84\%,  which is optimised using simulation to maximize the product of signal significance and signal efficiency. This criterion gives the optimal sensitivity to the oscillation frequency.

The remaining sources of background after selection consist of: random track combinations (combinatorial background); \BsDsstPi decays, where the photon from the $\Dssm\to\Dsm\gamma$ decay is not reconstructed; and contributions from \bquark-hadron decays with similar topologies to the signal, namely \BdDPi, \LbLcPi and \BsDsK decays. The decays with similar topology are suppressed by applying kinematic vetoes and additional particle identification requirements. 

\begin{figure}[t]
    \centering
    \includegraphics[width=\textwidth]{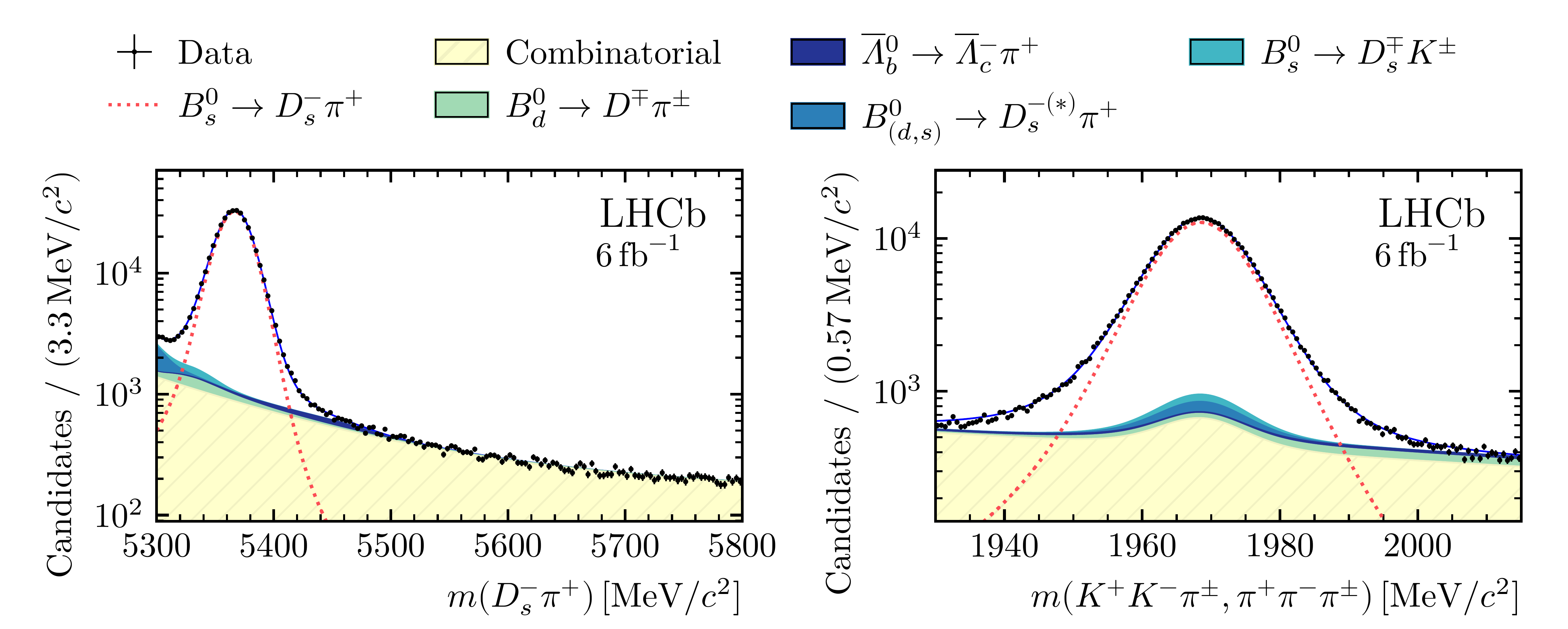}
    \caption{ 
        {\bf Invariant-mass distributions.}
        Distributions of the (left) $\Dsm\pip$, and (right)
        $\Kp\Km\pipm$ or $\pip\pim\pipm$ invariant
        mass for the selected candidates, $m(\Dsm\pip)$ and $m(\Kp\Km\pipm,\pip\pim\pipm)$, respectively. 
        The mass fit described in the text is overlaid. 
        The different contributions are shown as
        coloured areas (for background) or by dashed lines (for signal).
    The vertical bars, typically visible only in regions with low numbers of candidates, 
    correspond to the statistical uncertainty on the number of
    observed candidates in each bin. The horizontal bin width is indicated
    on the vertical axis legend.
    }
    \label{fig:massfit}
\end{figure}

In order to measure \dms, a \BsDsPi decay time distribution is first constructed in the absence of
background. This is achieved by performing an unbinned two-dimensional likelihood fit to the observed
$\Ds\pip$ and $\Km\Kp\pim$ or $\pim\pip\pim$ invariant-mass distributions.
This fit is used to determine the signal yield and a set of weights~\cite{Pivk:2004ty} used to
statistically subtract the background in the subsequent fit to the decay-time distribution.
The invariant mass distributions of the selected decays are shown in Fig.~\ref{fig:massfit}.
The non peaking contribution in the combinatorial background distribution, 
visible in Fig.~\ref{fig:massfit} (right), is due to events in which a fake \Ds candidate is produced from a combination of random tracks. The peaking contribution is due to genuine
$\Ds$ candidates combined with a random track resulting in a fake $\Bs$ candidate.

The probability density functions describing the signal and background invariant mass distributions
are obtained using a mixture of control samples in data and simulation.
The $\Dsm\pip$ and $\Km\Kp\pim$ or $\pim\pip\pim$ invariant-mass signal shapes are described by the
sum of a Hypatia~\cite{Santos:2013gra} and Johnson $S_U$~\cite{Johnson_dist} functions.
The combinatorial background contribution for both invariant-mass distributions is described
by an exponential function in each with parameters determined in the fit.
The \BdDPi, \LbLcPi or \BsDsK background components constitute less than 2\% of the signal yield 
and are accounted for in the fit to the invariant mass distributions using yields obtained from 
known branching fractions and relative efficiencies, as determined from simulated samples, which 
are weighted to account for differences between data and simulation.
 The \BdDsPi and \BsDsstPi background components are also obtained from simulated samples and included
 in the mass fit. The combined \BdDsPi and \BsDsstPi yield is a free parameter of the fit. 
 The signal yield obtained from the invariant mass fit is $378\,700\pm 700$. 

The decay-time parametrisation in Eq.~\ref{eq:time_eq} is modified to account for the 
following detector effects: 
a decay-time-dependent reconstruction efficiency; a time-dependent decay-time resolution; the imperfect knowledge of the initial flavour of the reconstructed \Bs or \Bsb meson; the asymmetry in \Bs or \Bsb production in $pp$ collisions; and an asymmetry in reconstruction of final state particles due to interactions in the detector material\cite{LHCb-PAPER-2016-062}. 

Due to the lifetime biasing effect of the selections, the reconstruction efficiency is low at small decay 
times and increases to a plateau after 2\ps.
The decay-time-dependent reconstruction efficiency is modelled with cubic b-splines curves as described in
Ref.~\cite{Karbach:ComplxErrFunc}.
The spline coefficients are allowed to vary in the fit to the observed decay-time distribution.

The decay-time resolution is measured using a data sample of \Dsm mesons originating from $pp$ interactions without being required to come from an intermediate \Bs meson decay. These `prompt' candidates pass the same real-time selection procedure as for the signal sample.
After real-time selection, additional requirements are applied to ensure a \Dsm signal peak with high background rejection 
but without any requirement on  displacement from the $pp$ collision point. The multivariate classifier trained
using the full signal decay is not applied. 
The reconstructed decay time in this control sample is proportional to the distance
between the \Dsm production vertex and an artificial \Bs decay vertex, formed by
combining the prompt \Dsm meson with a \pip track from the same $pp$ collision. It is therefore compatible with zero decay time up to bias and resolution effects. 
A linear relationship is observed between the decay-time resolution measured
at zero decay time and the decay-time uncertainty estimated in the vertex fit. 
This relationship is used to calibrate the \BsDsPi decay-time uncertainty.
Simulated prompt \Dsm and \BsDsPi decays, for which the generated decay time 
is known, are used to check the suitability of this method, which determines a 
$0.005 \ps$ bias in the reconstructed decay
time due to residual detector misalignments. This
bias is corrected for in the analysis. 
The uncertainty on \dms due to these residual detector misalignments, is evaluated using simulated samples that were intentionally
misaligned. This uncertainty is reported in Table~\ref{syst:tabsyst}.

To determine if a neutral meson oscillated into its antiparticle,
knowledge of the \Bs or \Bsb flavour at production and decay is required. In \BsDsPi decays, 
the \Bs flavour at decay is identified by the charge of the pion as the $\Dsp\pim$ decay cannot be produced directly. To determine whether the \Bs oscillated before decay,
the flavour at production is inferred from the hadronisation of the \Bs meson or 
the decay of other beauty hadrons produced in the collision
using a combination of several flavour-tagging 
algorithms~\cite{Fazzini:2018dyq,LHCb-PAPER-2011-027,LHCb-PAPER-2015-027,LHCb-PAPER-2015-056}.
Each algorithm estimates the probability that a candidate has been assigned the wrong flavour tag. The algorithms that use information independent of signal fragmentation are calibrated using \Bp meson decays and a combined wrong-tag estimate is used in the fit. The tagging efficiency is measured to be $\varepsilon=(80.30\pm0.07)\%$ with a
probability to tag a candidate as the wrong flavour of
$\omega=(36.21\pm0.17)\%$, where the uncertainties account for the calibration.

In the unbinned maximum likelihood fit to the decay-time distribution used to extract \dms, the calibration parameters for the combined wrong tag estimate are allowed to vary. Additional free parameters are the values of the spline coefficients used to describe
the decay-time-dependent reconstruction efficiency and the $\Bs$--$\Bsb$ production and detection asymmetries.

\begin{figure}
    \centering
    \includegraphics[width=0.49\textwidth]{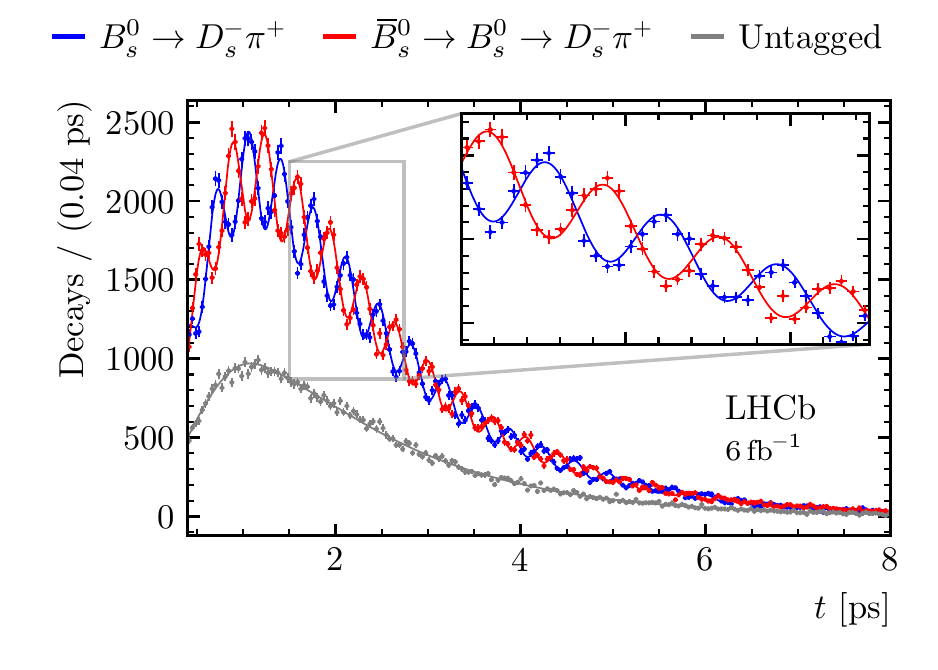}
    \includegraphics[width=0.49\textwidth]{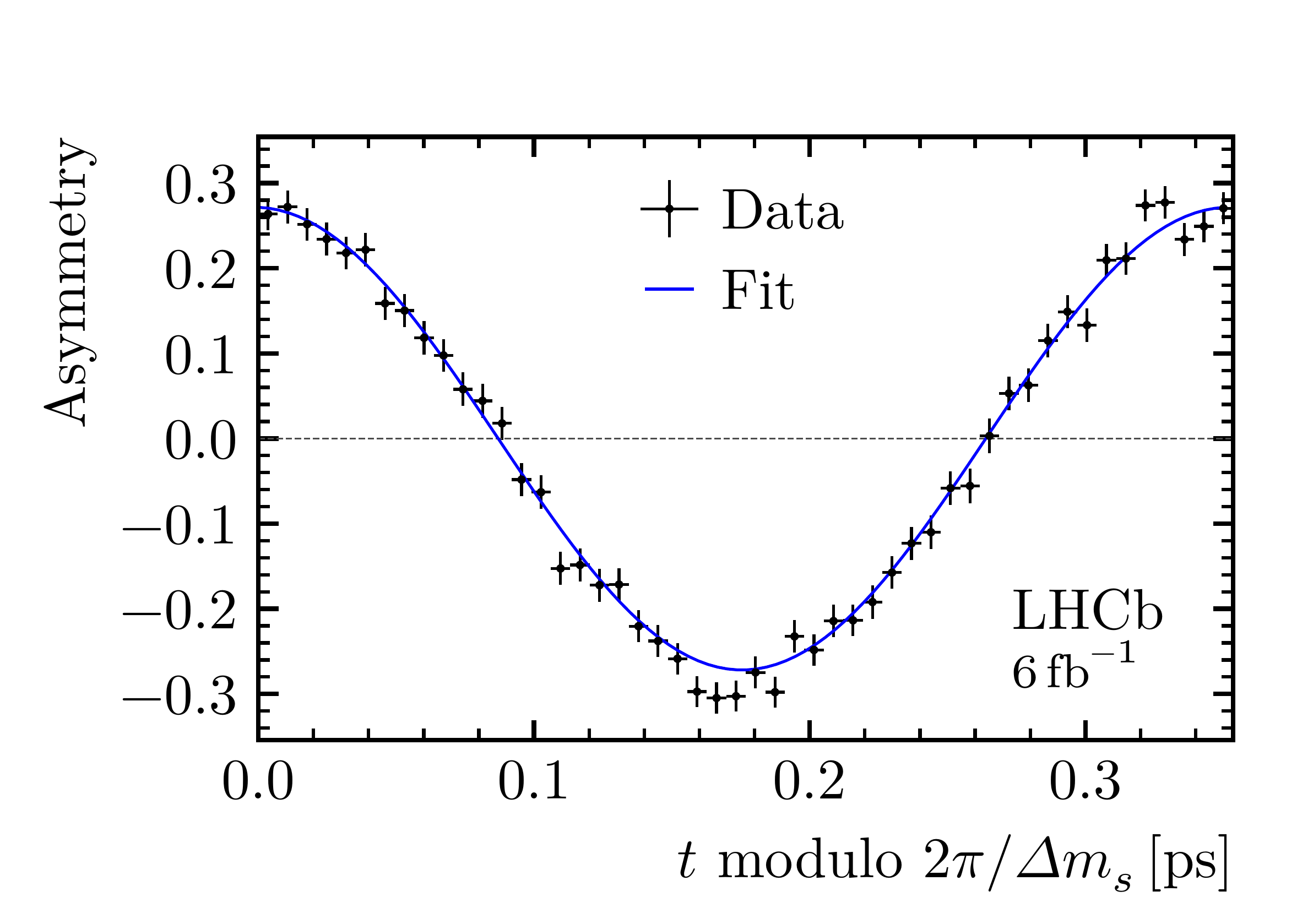}
    \caption{ 
    {\bf Decay-time distribution of the signal decays.}
    Distribution of the (left) decay time of the \BsDsPi signal decays and (right)
    decay-time asymmetry between mixed and unmixed signal decays.
    The vertical bars correspond to the statistical uncertainty 
    on the number of observed candidates in each bin. The horizontal bars
    represent the bin width. In the left plot, the horizontal bin width is indicated on the vertical axis legend. The three components, 
    unmixed, mixed and untagged, are shown in blue, red and gray, respectively. The insert
    corresponds to a zoom of the region delineated in grey.
    The fit described in the text is overlaid.}
    \label{fig:decaytimefit}
\end{figure}

The parameters $\Gamma_s$ and $\Delta\Gamma_s$, are fixed in the fit to their known values~\cite{PDG2020}. Other fixed parameters are: the estimate of the wrong-tag fraction and efficiency of the tagging algorithms, the decay-time bias correction and the 
decay-time resolution calibration parameters.
The decay-time distribution of the tagged--mixed, \BsbDsPi, tagged--unmixed, \BsDsPi, and untagged, where the initial flavour is unknown, samples
are shown in  Fig.~\ref{fig:decaytimefit} (left).
The corresponding fit projection is overlaid.
In order to highlight the oscillation phenomenon, the asymmetry
distribution between the tagged--unmixed and tagged--mixed samples is defined as
\begin{equation}
A(t) = \frac{N(\BsDsPi,t) - N(\BsbDsPi,t)}{N(\BsDsPi,t) + N(\BsbDsPi,t)},
\end{equation}
with $t$ modulo $2 \pi / \Delta m_s$, and is shown in Fig.~\ref{fig:decaytimefit} (right).
Here, $N(\BsbDsPi,t)$  and $N(\BsDsPi,t)$ indicate respectively the
tagged--mixed and tagged--unmixed decays observed at a time $t$.
For this distribution each event, in addition to the weight used to statistically subtract the 
background, is also weighted by the product of two factors. The
first is a flavour-tagging dilution factor, related to the probability that the
flavour tag is indeed correct. The second is an effective decay-time uncertainty
dilution factor, depending on the reconstructed decay time per-event resolution 
and on \dms, for which the central value of the decay time fit is being used.
The continuous line overlaid corresponds to the fit result. The result of the fit for $\dms$ is
$\resultstat\ps^{-1}$, where this uncertainty is related to the sample size.

Several sources of systematic uncertainty have been investigated and those with a non-negligible contribution are listed in Table~\ref{syst:tabsyst}.  These include the uncertainty on the momentum scale of the detector, obtained by comparing the reconstructed masses of known particles with the most accurate available values \cite{PDG2020}; residual detector misalignment and length scale uncertainties; and uncertainties due to the choice of mass and decay-time fit models, determined using alternate parametrisations and pseudoexperiments.
To verify the robustness of the measurement to variations in \dms as a function of the decay kinematics, the data sample is split into mutually disjoint subsamples, each having
 the same statistical significance, in relevant 
kinematic quantities, such as the \Bs momentum, and the \dms values obtained from each subsample are compared.
The largest observed variation is included as a systematic uncertainty.
The total systematic uncertainty is $\resultsyst\ps^{-1}$, with the leading contribution due to residual detector misalignment and detector length scale uncertainties.

\begin{table}[]
\centering
\caption{{\bf Systematic uncertainties affecting the measurement of \boldmath{\dms}.} Sources of 
systematic uncertainties are discussed in  the text. Additional details are provided in the Methods
section. The total systematic uncertainty is obtained by adding the contributions in
quadrature.}
\label{syst:tabsyst}
\begin{tabular}{lc}
Description & Systematic uncertainty [$\ps^{-1}$] \\ \hline
Reconstruction effects:                            &        \\ 
        \quad momentum scale uncertainty           & 0.0007 \\
         \quad detector length scale               & 0.0018 \\
         \quad detector misalignment               & 0.0020 \\
Invariant mass fit model:                          &        \\ 
         \quad background parametrisation          & 0.0002 \\
         \quad  $\BsDsstPi$  and   $\BdDsPi$  contributions & 0.0005 \\ 
Decay-time fit model:                              &        \\

        \quad decay-time resolution model          & 0.0011 \\

        \quad neglecting correlation among observables  & 0.0011     \\ 
Cross-checks:                                      &        \\ 
        \quad kinematic correlations               & 0.0003 \\ \hline
Total systematic uncertainty                       & {\resultsyst}   \\ \hline

\end{tabular}
\end{table}

\begin{figure}
    \centering
    \includegraphics[width=1.0\textwidth]{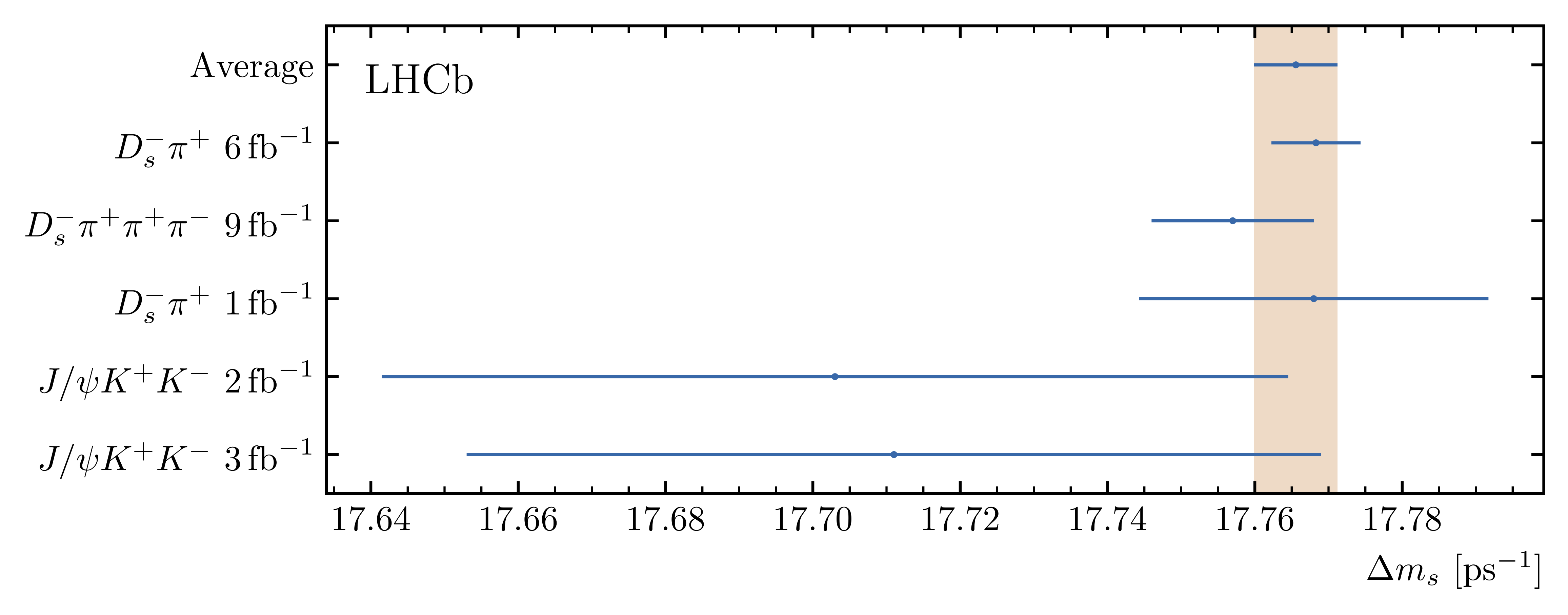}
    \caption{ 
    {\bf Summary of LHCb measurements.}
    Comparison of LHCb $\dms$ measurements from
    Refs.~\cite{LHCb-PAPER-2013-006,LHCb-PAPER-2020-030,LHCb-PAPER-2014-059,LHCb-PAPER-2019-013}, 
    the result presented in this article and their average. The measurement described in this paper is 
    labeled as $\Dsm \pi^+$ 6\invfb. The horizontal bars correspond to the total uncertainty reported for each measurement.
    The band indicates the size of the uncertainty on the average for
    comparison purposes. The combination procedure and inputs are described in the text.}
    \label{fig:combination}
\end{figure}

The value of the \Bs--\Bsb oscillation frequency determined in this article:
\begin{equation}
    \dms = \result \ps^{-1} \nonumber
\end{equation}
is the most precise measurement to date. 
The precision is further enhanced by combining this result with the values determined in Refs.~\cite{LHCb-PAPER-2013-006,LHCb-PAPER-2020-030}. Reference~\cite{LHCb-PAPER-2013-006} 
uses \BsDsPi decays collected in 2011. Reference~\cite{LHCb-PAPER-2020-030} uses a sample of 
$\Bs \to \Ds \pip \pip \pim$ decays selected from the combined 2011--2018 data set, corresponding to 9\invfb.
The measurements are statistically independent. 
The systematic uncertainties related to the momentum scale, length scale 
and residual detector misalignment are assumed to be fully correlated. 
Due to aging of the detector and different alignment procedures used in Run 1 and Run 2, the effect of residual detector misalignment is larger in measurements using Run 2 data. Given the precision of the measurement described in this paper, a detailed study of the detector misalignment effects is performed and the related uncertainty due to the decay time bias has been reduced significantly compared to previous measurements using the Run 2 data.
The values of the fixed parameters $\Delta\Gamma_s$ and $\Gamma_s$ used as inputs to the previous analyses 
have evolved over time as additional measurements have been made. However as
the correlation between \dms and $\Delta\Gamma_s$ and $\Gamma_s$ is negligible
these small differences have been ignored in the combination procedure. 
A covariance matrix is constructed by adding statistical and systematic uncertainties in quadrature for each 
input, including correlations between systematic uncertainties. 
The results are averaged by minimizing the $\chi^2$ from the full covariance matrix. 
The value of $\dms$ obtained is $\combinationHadronic\ps^{-1}$. 
Additionally, these results are combined with those from Refs.~\cite{LHCb-PAPER-2014-059,LHCb-PAPER-2019-013} 
where \dms is determined using $\Bs \rightarrow \jpsi K^+ K^-$ decays in the 2011--2012 (3\invfb) and
2015--2016 (2\invfb) data sets, respectively.
The decay-time resolution for the measurements used in the combination, see Refs.~\cite{LHCb-PAPER-2013-006,LHCb-PAPER-2014-059,LHCb-PAPER-2019-013,LHCb-PAPER-2020-030}, including the analysis presented here, varies from 35 to 45 fs, depending on the decay mode.
The result for \dms is  
 $\combinationAll\ps^{-1}$. 
 The different measurements, and the resulting combination, are shown in Fig.~\ref{fig:combination}.

In summary, this paper presents the most precise measurement of the \dms oscillation frequency, 
$\result \ps^{-1}$,
where the first uncertainty is statistical and the second systematic. The result is obtained using a sample of \BsDsPi
decays collected with the LHCb detector during Run 2 of the LHC.
Combining the result of this paper with previous measurements by the LHCb collaboration yields a \dms value of
$\combinationAll\ps^{-1}$.
This value is compatible with, and considerably more precise than, the predicted value from lattice 
QCD \cite{Phys.Rev.D93.2016.113016,1812.08791,1907.01025}
and sum rule calculations \cite{1806.00253,1904.00940}
of $18.4^{+0.7}_{-1.2}\ps^{-1}$\cite{DeltaMsTheoUpdate}. The combined result represents a significant
improvement over previous measurements, and is a legacy measurement of the original LHCb detector. 
The experiment is currently undergoing a major upgrade to operate at five times the instantaneous luminosity
from 2022 onwards\cite{LHCb-TDR-012}. 
The largest sources of systematic uncertainty for this measurement, namely those related to the detector
length scale and misalignment, will 
be a focal point to further improve upon this result for future data taking periods.

\section*{Methods}

\noindent {\bf The LHCb detector.}
 The \lhcb detector~\cite{LHCb-DP-2008-001,LHCb-DP-2014-002} is a single-arm forward
spectrometer covering the \mbox{pseudorapidity} range $2<\eta <5$,
designed for the study of particles containing \bquark or \cquark
quarks. The detector includes a high-precision tracking system
consisting of a silicon-strip vertex detector surrounding the $pp$
interaction region~\cite{LHCb-DP-2014-001}, a large-area silicon-strip detector located
upstream of a dipole magnet with a bending power of about
$4{\mathrm{\,Tm}}$, and three stations of silicon-strip detectors and straw
drift tubes~\cite{LHCb-DP-2017-001} placed downstream of the magnet.
The tracking system provides a measurement of the momentum, \ptot, of charged particles with
a relative uncertainty that varies from 0.5\% at low momentum to 1.0\% at 200\gevc.
The minimum distance of a track to a primary vertex (PV), the impact parameter (IP), 
is measured with a resolution of $(15+29/\pt)\mum$,
where \pt is the component of the momentum transverse to the beam, in\,\gevc.
Different types of charged hadrons are distinguished using information
from two ring-imaging Cherenkov detectors~\cite{LHCb-DP-2012-003}. 
Photons, electrons and hadrons are identified by a calorimeter system consisting of
scintillating-pad and preshower detectors, an electromagnetic
and a hadronic calorimeter. Muons are identified by a
system composed of alternating layers of iron and multiwire
proportional chambers~\cite{LHCb-DP-2012-002}.

Simulation of the \lhcb detector response is required to model the effects of the detector acceptance 
and the imposed selection requirements.
In the simulation, $pp$ collisions are generated using \pythia~\cite{Sjostrand:2007gs} with a specific 
\lhcb configuration~\cite{LHCb-PROC-2010-056}.
Decays of unstable particles are described by \evtgen~\cite{Lange:2001uf}, in which final-state
radiation is generated using \photos~\cite{davidson2015photos}.
The interaction of the generated particles with the detector, and its response, are implemented using 
the \geant toolkit~\cite{Allison:2006ve, *Agostinelli:2002hh} as described in
Ref.~\cite{LHCb-PROC-2011-006}.

\noindent {\bf Selection.} A fast decision about which $pp$ collisions are of interest is made by a 
trigger system~\cite{LHCb-DP-2012-004}.
It consists of a hardware stage, based on information from the calorimeter and muon systems, followed by
a software stage, which  reconstructs the $pp$ collision based on all available detector information.
The software trigger selects candidates consistent with a \bquark-hadron decay topology, with tracks
originating from a vertex detached from the primary $pp$ collision point, known as the primary vertex (PV).
The mean \Bs lifetime is 1.5\ps~\cite{PDG2020}, which corresponds to an average flight distance of 1\cm in the
\lhcb detector.

After being accepted by the trigger, a further selection is applied which forms \DsKKPi and \DsPiPiPi 
candidates from reconstructed charged tracks and subsequently combines them with a fourth
track to form \BsDsPi candidates.
Particle identification information is used to assign mass hypotheses to each of the final-state tracks.

The \Bs invariant-mass resolution is improved by constraining the \Ds invariant 
mass to its known value~\cite{PDG2020}.
The $\Kp\Km\pipm$ or $\pip\pim\pipm$ and $\Dsm\pip$ invariant-mass ranges considered in this analysis are 
[1930,2015] and $[5300,5800] \mevcc$, respectively.

To suppress \Bs candidates formed from random track combinations, a gradient boosted decision tree (BDT)
is used, implemented in the XGBoost library \cite{XGBoost}.
The training uses data for both the signal and the background samples in order to avoid mismatches 
between data and simulation. 
This classifier uses information on: the fit quality of the $\Ds$ and $\Bs$ decay vertices; the $\Ds$ and $\Bs$ $\chi^2_{\rm IP}$ defined as the difference in the $\chi^2$ of the vertex fit for a given PV reconstructed with and without the considered particle; the angles between their momentum vector and the vector connecting their production and decay vertices; and the $\pt$ and impact parameter $\chi^2_{\rm IP}$ of the final-state tracks. 
The BDT classifier threshold is chosen to maximize the product of the signal significance
and the signal efficiency. This choice optimises sensitivity to the oscillation frequency.

\noindent {\bf Flavour tagging.} 
The initial flavour of the \Bs meson must be known in order to determine if it has oscillated prior to decay.
Flavour tagging algorithms are used to determine the initial flavour from properties of the \bquark-hadron production in the $pp$ collision.

Beauty quarks are predominantly produced in pairs. Opposite side (OS) tagging algorithms~\cite{LHCb-PAPER-2011-027} determine the initial flavour of the \Bs meson based on information from the other beauty-quark decay. These include the
OS muon and OS electron taggers, which identify the flavour from the charge of leptons produced in the other $b$-hadron decay.
The OS kaon tagger identifies $b \rightarrow c \rightarrow s$ transitions, the OS
charm quark tagger identifies $b \rightarrow c$ transitions, and the OS vertex
charge tagger calculates the effective charge of an OS displaced
vertex~\cite{LHCb-PAPER-2015-027}.
In addition, a same side (SS) kaon tagger exploits the charge information of
the kaon originating from the $\bar{s}$ or $s$ quark leftover from the \Bs or \Bsb meson
fragmentation~\cite{LHCb-PAPER-2015-056}.
Each algorithm determines the initial flavour of the \Bs meson from the
charge of the reconstructed tagging particle or the reconstructed vertex in the case of the OS vertex tagger.

The tagging information is incorporated in the decay-time description.
The amplitude of the oscillation is reduced by a dilution factor $D=(1-2\omega)$, with $\omega$ the
average fraction  of incorrect tags known as the mistag rate in the literature.
Different machine learning algorithms provide an estimate of the mistag rate which is calibrated with data to match the true mistag distribution.
A linear calibration of the average mistag as a function of the predicted mistag for the combined OS tag and SS
kaon tag information is then implemented in the decay-time fit with freely varying calibration
parameters. 
The combined tagging efficiency of the sample is $\varepsilon=(80.30\pm0.07)\%$ with a mistag fraction of $\omega=(36.21\pm0.02\pm0.17)\%$ where the first uncertainty is due to the finite size of the calibration sample and the second is due to the calibration procedure. This results in a combined effective performance of $(6.10\pm0.02\pm0.15)\%$ with respect to a perfect tagging algorithm which would have a 100\% tagging efficiency and zero mistag rate.

\noindent {\bf Decay time fit.}
The observed decay-time distribution is fitted using an 
unbinned maximum likelihood fit in which all combinations of initial
state flavours (\Bs, \Bsb, or untagged) and final state charges (\Dsm\pip or \Dsp\pim) are fit simultaneously.
The decay-time distribution of each measured final state is described by the sum of all
processes contributing to that state.
Experimental effects are taken into account with several adjustments to the
theoretical prediction in eq.~\ref{eq:time_eq}, namely:
\begin{align}
\label{eq:time_eq_exp}
    P_\mathrm{exp.}(\BsDsPi, t) \sim  (1 + a_\mathrm{det}) \cdot &  \big[ \nonumber \\ 
    & (1 - \omega) (1 - a_\mathrm{prod}) P(\BsDsPi, t) \nonumber \\  
    & + \omega (1 + a_\mathrm{prod}) P(\BsbDsPi, t) \big]\,.
\end{align}
Production and detection asymmetries are parameterised by
factors $a_\mathrm{prod}$ and $a_\mathrm{det}$, respectively, which are allowed to deviate from unity.
The decay-time distribution of both flavours contain a fraction $1 - \omega$ of the correctly tagged decay-time parametrisation plus a
fraction $\omega$ of the incorrectly tagged decay-time parametrisation.
The mistag rate $\omega$ is  obtained from a per-event estimation, after a linear calibration. Different calibration parameters are used
for the \Bs and \Bsb initial flavours.

The experimental decay-time distributions of both flavours are convolved with a Gaussian function to account for the finite detector resolution.
The mean of this function is shifted by the decay-time bias correction factor,
and the width is obtained from a per-event estimate of the decay-time uncertainty after a linear
calibration.

A decay-time dependent efficiency is finally modelled by a time
dependent cubic spline function, which multiplies the decay-time distribution obtained from
the previous step.

\noindent {\bf Systematic uncertainties.} The following sources of systematic
uncertainty have been found to give a non negligible contribution to the $\dms$
measurement. These are summarised in Table~\ref{syst:tabsyst}. 
 
 The measured decay-time is inversely proportional to the $\Bs$ momentum, and therefore depends upon an accurate determination of the momentum scale uncertainty of the tracking system. The uncertainty is determined by varying the \Bs meson momentum by $\pm 0.03 \%$ (coming from a comparison of masses of different particles with their known values) in simulated signal samples. The corresponding uncertainty on \dms is $0.0007 \ps^{-1}$.

The measured decay time is also proportional to the distance the \Bs meson travels between production and decay, which is affected by precise knowledge of the position of the vertex detector elements along the proton beam axis. The measured uncertainty is 100\mum over a length of 1\m~\cite{LHCb-DP-2014-001}.  The corresponding uncertainty on \dms is $0.0018 \ps^{-1}$.

The relative alignment of the tracking detector elements are a source of bias in the decay-time and contribute to resolution effects.
The uncertainty on \dms due to imprecise knowledge of this alignment has been obtained from the analysis of 
simulated signal samples in which the detector elements have been deliberately misaligned.
Different misalignments, translations, rotations and combinations of both, have been investigated.
The leading effect is due to translation along the $x$-axis, the axis perpendicular to the beam direction pointing towards the center of the LHC ring.
As a consequence simulated signal samples have been misaligned with $x$-axis translations
in the range between 0 and 9 \mum as determined from survey results. 
Each misaligned simulated sample is then corrected for decay time bias in the same manner as for data,
and the extracted \dms value is compared with the value obtained in simulation without any
misalignment.
This comparison produces a corresponding uncertainty on the bias correction procedure of 0.0020 $\ps^{-1}$.

    Alternative parametrisations of the background contributions to the invariant mass fit have 
    been obtained by using different weighting methods; the 
    difference between these parametrisations corresponds to an uncertainty of $0.0002 \ps^{-1}$.
    
    For the specific $\BsDsstPi$ and $\BdDsPi$ background contributions, the relative fraction of these 
    components cannot be reliably determined from the data. Their relative contributions are nominally set to an equal mixture and varied between 0 (pure $\BdDsPi$) and 1 (pure $\BsDsstPi$) to determine the maximum deviation in \dms corresponding to an uncertainty of $0.0005 \ps^{-1}$.    
    
    The decay--time resolution is obtained from data using a sample of \Dsm mesons that are produced directly in $pp$ collision. These are combined with a \pip meson coming from the same collision to produce a fake \Bs candidate with a decay time equal to zero, ignoring resolution effects. Different parametrisations of the measured decay-time distribution are applied to a simulated signal sample.  The largest deviation of the extracted \dms value with respect to the nominal parametrisation is found to be $0.0011 \ps^{-1}$.
    
    The procedure used to subtract background contributions in the fit to the decay-time distribution assumes no large correlations between the decay-time and the reconstructed \Bs and \Ds invariant masses. This is studied by analysing 
    simulated signal and background samples where any residual correlations between these observables are removed. The difference in measured value of \dms between the decorrelated and nominal samples is found to be $0.0011 \ps^{-1}$.
    
    The data sample was split into mutually disjoint subsamples in order to study the effect of potential correlations between kinematic ranges, data taking periods, flavour-tagging categories, the BDT-based selection and the measured value of \dms. The measured values obtained from each subsample are compared and the largest observed variation is found to be $0.0003 \ps^{-1}$.  

Several additional effects have been considered consisting of: possible biases introduced by the fit procedure,
changes to the signal and background parametrisations, and changes in the reweighting procedure used when obtaining the invariant mass shapes of partially reconstructed backgrounds constituting less than 2 \% of the signal yield. Their impact has been found to be negligible with respect to the sources listed in Table~\ref{syst:tabsyst}.

The largest sources of systematic uncertainty are found to be due to imprecise knowledge of the  
position and alignment of the tracking detector closest to the nominal $pp$ collision region.

\vspace{1cm}
\noindent {\bf Data Availability.}
All figures are available from: \\
\href{https://lhcbproject.web.cern.ch/Publications/p/LHCb-PAPER-2021-005.html}{https://lhcbproject.web.cern.ch/Publications/p/LHCb-PAPER-2021-005.html} \\
Additional material describing this analysis is available from: \\
http://cds.cern.ch/record/2764338/files/ \\
in LHCb-PAPER-2021-005-supplementary.zip as Supplementary.pdf \\
Inputs to the $\Delta m_s$ combination are
available from HEPdata at: \\
https://www.hepdata.net/record/105881 \\
LHCb has an open data policy described in: \\ 
\href{http://cdsweb.cern.ch/record/1543410?ln=en}{http://cdsweb.cern.ch/record/1543410} \\
Subject to the resources being identified, LHCb will endeavor to provide open 
access to some reconstructed level data on disk at CERN.

\vspace{1cm}

\noindent {\bf Code Availability.} The code used for the fit to the \Bs and \Ds invariant 
mass distributions and to the \Bs decay time distribution is publicly available at: \\
\href{https://gitlab.cern.ch/lhcb/Urania/-/tree/master/PhysFit/B2DXFitters}
{https://gitlab.cern.ch/lhcb/Urania/-/tree/master/PhysFit/B2DXFitters}

\section*{Acknowledgements}
%
%
\noindent We express our gratitude to our colleagues in the CERN
accelerator departments for the excellent performance of the LHC. We
thank the technical and administrative staff at the LHCb
institutes.
We acknowledge support from CERN and from the national agencies:
CAPES, CNPq, FAPERJ and FINEP (Brazil); 
MOST and NSFC (China); 
CNRS/IN2P3 (France); 
BMBF, DFG and MPG (Germany); 
INFN (Italy); 
NWO (Netherlands); 
MNiSW and NCN (Poland); 
MEN/IFA (Romania); 
MSHE (Russia); 
MICINN (Spain); 
SNSF and SER (Switzerland); 
NASU (Ukraine); 
STFC (United Kingdom); 
DOE NP and NSF (USA).
We acknowledge the computing resources that are provided by CERN, IN2P3
(France), KIT and DESY (Germany), INFN (Italy), SURF (Netherlands),
PIC (Spain), GridPP (United Kingdom), RRCKI and Yandex
LLC (Russia), CSCS (Switzerland), IFIN-HH (Romania), CBPF (Brazil),
PL-GRID (Poland) and NERSC (USA).
We are indebted to the communities behind the multiple open-source
software packages on which we depend.
Individual groups or members have received support from
ARC and ARDC (Australia);
AvH Foundation (Germany);
EPLANET, Marie Sk\l{}odowska-Curie Actions and ERC (European Union);
A*MIDEX, ANR, Labex P2IO and OCEVU, and R\'{e}gion Auvergne-Rh\^{o}ne-Alpes (France);
Key Research Program of Frontier Sciences of CAS, CAS PIFI, CAS CCEPP, 
Fundamental Research Funds for the Central Universities, 
and Sci. \& Tech. Program of Guangzhou (China);
RFBR, RSF and Yandex LLC (Russia);
GVA, XuntaGal and GENCAT (Spain);
the Leverhulme Trust, the Royal Society
 and UKRI (United Kingdom).

\section*{Author contributions}
All authors have contributed to the publication, being variously involved in the
design and the construction of the detectors, in writing software, in operating 
the detectors and acquiring data, in calibrating sub-systems and processing data
and finally in analysing the processed data.

\section*{Competing interests}
The authors declare no competing interests.



\addcontentsline{toc}{section}{References}
\bibliographystyle{LHCb}
\bibliography{main,standard,LHCb-PAPER,LHCb-CONF,LHCb-DP,LHCb-TDR}

\newpage
\centerline
{\large\bf LHCb collaboration}
\begin
{flushleft}
\small
R.~Aaij$^{32}$,
C.~Abell{\'a}n~Beteta$^{50}$,
T.~Ackernley$^{60}$,
B.~Adeva$^{46}$,
M.~Adinolfi$^{54}$,
H.~Afsharnia$^{9}$,
C.A.~Aidala$^{86}$,
S.~Aiola$^{25}$,
Z.~Ajaltouni$^{9}$,
S.~Akar$^{65}$
% ...

{\footnotesize \it

% ...
$^{9}$Universit{\'e} Clermont Auvergne, CNRS/IN2P3, LPC, Clermont-Ferrand, France\\
% ...
$^{25}$INFN Sezione di Milano, Milano, Italy\\
% ...
$^{32}$Nikhef National Institute for Subatomic Physics, Amsterdam, Netherlands\\
% ...
$^{46}$Instituto Galego de F{\'\i}sica de Altas Enerx{\'\i}as (IGFAE), Universidade de Santiago de Compostela, Santiago de Compostela, Spain\\
% ...
$^{50}$Physik-Institut, Universit{\"a}t Z{\"u}rich, Z{\"u}rich, Switzerland\\
% ...
$^{54}$H.H. Wills Physics Laboratory, University of Bristol, Bristol, United Kingdom\\
% ...
$^{60}$Oliver Lodge Laboratory, University of Liverpool, Liverpool, United Kingdom\\
% ...
$^{65}$University of Cincinnati, Cincinnati, OH, United States\\
% ...
$^{68}$Syracuse University, Syracuse, NY, United States\\
% ...
$^{86}$University of Michigan, Ann Arbor, United States, associated to $^{68}$\\
\bigskip
% ...
\medskip
}
\end{flushleft}

\end{document}